\documentclass[12pt,a4paper]{iopart}
\usepackage{graphicx}
\usepackage{color}
\newcommand{\be}{\begin{equation}}
\newcommand{\ee}{\end{equation}}
\newcommand{\bea}{\begin{eqnarray}}
\newcommand{\eea}{\end{eqnarray}}
\begin{document}
\title[Energy density at kinetic freeze-out]{Energy density at kinetic freeze-out in Pb-Pb collisions at the LHC using the Tsallis distribution}
\author{M.~D. Azmi$^1$, T. Bhattacharyya$^2$, J. Cleymans$^3$\\ and M. Paradza$^3$}
\address{$^1$ Physics Department, Aligarh Muslim University, Aligarh-202002 (U.P.), India}
\address{$^2$ Bogoliubov Laboratory of Theoretical Physics, JINR, Dubna, 141980, Moscow region, Russia}
\address{$^3$UCT-CERN Research Centre and Department of Physics, University of Cape Town, 
Rondebosch 7701, South Africa}
\date{\today}
%
%
\begin{abstract}
The thermodynamic parameters like energy density, pressure, entropy density, temperature and particle density
are determined from the transverse momentum distributions of charged hadrons in Pb-Pb collisions at the LHC. 
The results show a clear increase with the centrality and the beam energy in all parameters. 
It is determined that in the final freeze-out stage the energy density
reaches a value of about 0.039 GeV/fm$^3$ for the most central collisions at $\sqrt{s_{NN}}$ =  5.02 TeV. This is less than that at chemical
freeze-out 
where the energy density is about 0.36 GeV/fm$^3$. This decrease approximately follows a $T^4$ law.
The results for the pressure and entropy density are also presented for each centrality class at $\sqrt{s_{NN}}$ =
 2.76 and 5.02 TeV.

\end{abstract}
%
%
\section{\label{secIntroduction}Introduction}

In heavy-ion collisions at the Large Hadron Collider (LHC) quark-gluon matter is created at a very  high energy density.
After the initial very hot stage, the system expands, reaches  chemical equilibrium and then finally  freezes-out in a  stage usually referred
to as the kinetic freeze-out stage.  
The present paper  determines the energy density, the pressure, the entropy density and the particle density
at this final kinetic freeze-out stage using the transverse momentum
distributions of  charged hadrons  measured  by the  ALICE collaboration in Pb-Pb collisions at $\sqrt{s_{NN}}$ = 2.76 and 5.02 TeV~\cite{Acharya:2018qsh}.  
For this purpose one needs a description which is thermodynamically consistent, i.e. the following relations must be satisfied:
\begin{eqnarray}
d\epsilon=T~ds +\mu~dn,\label{eq1}\\
dP=s~dT + n~d\mu\label{eq2},
\end{eqnarray}
where $\epsilon$ is the energy density, $T$ is the temperature, $s$ is the entropy density, $P$ is the pressure, $\mu$ is the chemical potential and
$n$ is the particle density. The   Maxwell relations given below  follow from this:
\begin{eqnarray}
T=\left.\frac{\partial\epsilon}{\partial s}\right|_n,\label{temp}~~~~~~~\mu=\left.\frac{\partial\epsilon}{\partial n}\right|_s,\label{mu}\\
n=\left.\frac{\partial P}{\partial \mu}\right|_{T},\label{number}~~~~~~s=\left.\frac{\partial P}{\partial T}\right|_\mu.\label{ent}
\end{eqnarray}
The following thermodynamic relation must also be  satisfied:
\begin{equation}
\epsilon + P = T s + \mu n  .
\end{equation}
Such a description  does exist.  It is based on the Tsallis distribution~\cite{Tsallis:1987eu} given by:
\be 
f(E,q,T,\mu) \equiv \left(1+(q-1)\frac{E-\mu}{T}\right)^{-\frac{1}{q-1}},
\label{tsd1}
\ee
where $E$ is the energy of the particle,  $q$ is the Tsallis parameter which, when approaching 1, makes the 
function $f$ exponential (Boltzmann-like). 
The chemical potential $\mu$ will be taken to be zero in the present analysis presented below as is appropriate in the 
central rapidity region at LHC energies. 

There exist other, closely related, distributions that have been used to describe transverse momentum spectra, 
see~\cite{Abelev:2013ala,Biyajima:2006mv,Lao:2016gxv,Si:2017cyg,Biro:2017arf,Hui:2017zqy} but these do not offer 
the advantage of  connecting to a full thermodynamic
description.

The relevant thermodynamic quantities can be obtained from the following relations

\begin{eqnarray}
s &=& - g \int\frac{d^3p}{(2\pi)^3}\left(\frac{f - f^q}{1-q} -f\right),\label{entropy}\\
n &=& g \int\frac{d^3p}{(2\pi)^3} f^{q},\label{Number}\\
\epsilon &=& g \int\frac{d^3p}{(2\pi)^3}~E~f^{q},\label{epsilon}\\
P &=& g\int\frac{d^3p}{(2\pi)^3}\frac{p^{2}}{3E}~f^{q}\label{pressure},
\end{eqnarray}
where $g$ is the usual degeneracy factor.  It is to be emphasized that the variable $T$ 
appearing in equation (6) obeys the thermodynamic relation
\begin{equation}
T = \left.\frac{\partial E}{\partial S}\right|_{N,V} ,
\end{equation}
and hence, the parameter $T$ can be called a temperature, albeit for a system obeying Tsallis and not Boltzmann-Gibbs statistics.

The derivation has been presented in  detail in~\cite{Cleymans:2011in,Cleymans:2012ya}. 
It has also been shown that it leads to a surprisingly good fit of  transverse momentum 
distributions up to $p_T$ values of 200 GeV/c~\cite{Wong:2013sca,Wong:2015mba,Azmi:2014dwa,Azmi:2015xqa}. A  comprehensive
comparison with experimental results  has been presented in~\cite{Grigoryan:2017gcg} for pp collisions.

The picture of a  heavy-ion collision proposed is as follows: the chemical freeze-out happens 
with Boltzmann-Gibbs statistics leading to a consistent picture of the hadronic yields, 
kinetic freeze-out
follows, this time obeying Tsallis statistics as proposed in this and previous papers, at least for values of the transverse momentum
up to 3 GeV and possibly even far beyond.  Clearly the Tsallis distribution, being a polynomial one characterized
by the Tsallis parameter $q$, contains information about collective flow.

Our paper is organized as follows. In  section 2 we discuss the single particle distribution 
which we use for fitting the transverse momentum spectra of the charged hadrons, 
and determine the temperature $T$ and the Tsallis parameter $q$.  In section 3 we calculate
the corresponding 
thermodynamic quantities namely, the energy density, the pressure, the entropy density and the particle density. 
The values obtained for the energy density are then discussed and compared to values obtained at different stages
of the collision and to other closely related energy densities.
Lastly, we summarize our results and conclude in section 4.

\section{Transverse Momentum Distribution}
The Tsallis 
distribution was  first proposed more than three decades ago as a generalization of 
the  Boltzmann-Gibbs distribution~\cite{Tsallis:1987eu}, and is characterized by only 
three parameters namely, the Tsallis parameter $q$, the temperature  $T$ and  the volume $V$.
 
  The momentum distribution of particles obtained by using the expression for the particle density given in equation~\eref{Number} 
is written as:
\begin{equation}
\frac{d^3N}{d^3p}=\frac{gV}{(2\pi)^3}\left[1 + (q-1)\frac{E-\mu}{T}\right]^{-q/(q-1)}.
\end{equation}
When expressed in terms of transverse momentum, $p_{{T}}$, transverse mass, $m_{{T}} = \sqrt{{p_{{T}}}^2 + m^2}$ and
 rapidity, $y$, the above equation takes the following form:
\begin{equation}
\frac{d^2 N}{dp_{{T}} dy}=gV\frac{p_{{T}}~m_{{T}} \cosh y}{(2\pi)^2}
\left[1 + (q-1)\frac{m_{{T}}{\rm cosh}y - \mu}{T}\right]^{-q/(q-1)}.
\label{ptdist}
\end{equation}
\indent At mid-rapidity, $y$ = 0, and for zero chemical potential, as is relevant at the LHC,  equation~\eref{ptdist}
reduces to the following expression:
\begin{eqnarray}
\left.\frac{d^{2}N}{dp_{{T}}~dy}\right|_{y=0} = gV\frac{p_{{T}}m_{{T}}}{(2\pi)^2}\left[1+(q-1)\frac{m_{{T}}}{T}\right]^{-q/(q-1)}  .
\label{tsallisfit}
\end{eqnarray}

The transverse momentum distributions of charged particles produced in Pb-Pb collisions at LHC energies are fitted using a 
sum of three Tsallis distributions. These consist of fits for $\pi^+$'s, $K^+$'s and protons. 
The following expression, at mid-rapidity and $\mu = 0$, was used to fit the
pseudo-rapidity distributions: 
\begin{equation}
\frac{d^{2}N_{ch}}{dp_T~d\eta} = 2p_T\frac{V}{(2\pi)^2}
\sum_{i=1}^{3}g_im_{T,i}\frac{p_T}{m_{T,i}}\left[1+(q-1)\frac{m_{T,i}}{T}\right]^{\frac{-q}{(q-1)}}
\label{tsallisfitcharged}
\end{equation}
where $i = \pi^+, K^+, p$. 
The relative weights between particles are given by by the corresponding degeneracy factors 
 $g_{\pi^+}$ = $g_{K^+}$ = 1 and $g_p$ = 2.
The factor 2 on the right hand side takes into account the contributions from antiparticles, $\pi^-, K^-$ and $\bar{p}$.
The extra factor $p_T/m_T$ on the right hand side takes into account the change from rapidity to pseudo-rapidity
using the relation:
\begin{eqnarray}
\frac{dN}{dp_T~d\eta} = \sqrt{1 - \frac{m^2}{m_T^2 \cosh^2y}} \frac{dN}{dy dp_T} ,
\end{eqnarray}
which, at mid-rapidity, becomes:
\begin{equation}
\frac{dN}{dp_T~d\eta} = \frac{p_T}{m_T}\frac{dN}{dp_T~dy} ,
\end{equation}
hence the extra factor of $p_T/m_T$.

\begin{figure}[ht]
\begin{minipage}{\columnwidth}
\centering
\includegraphics[width=\columnwidth, height = 12.0cm]{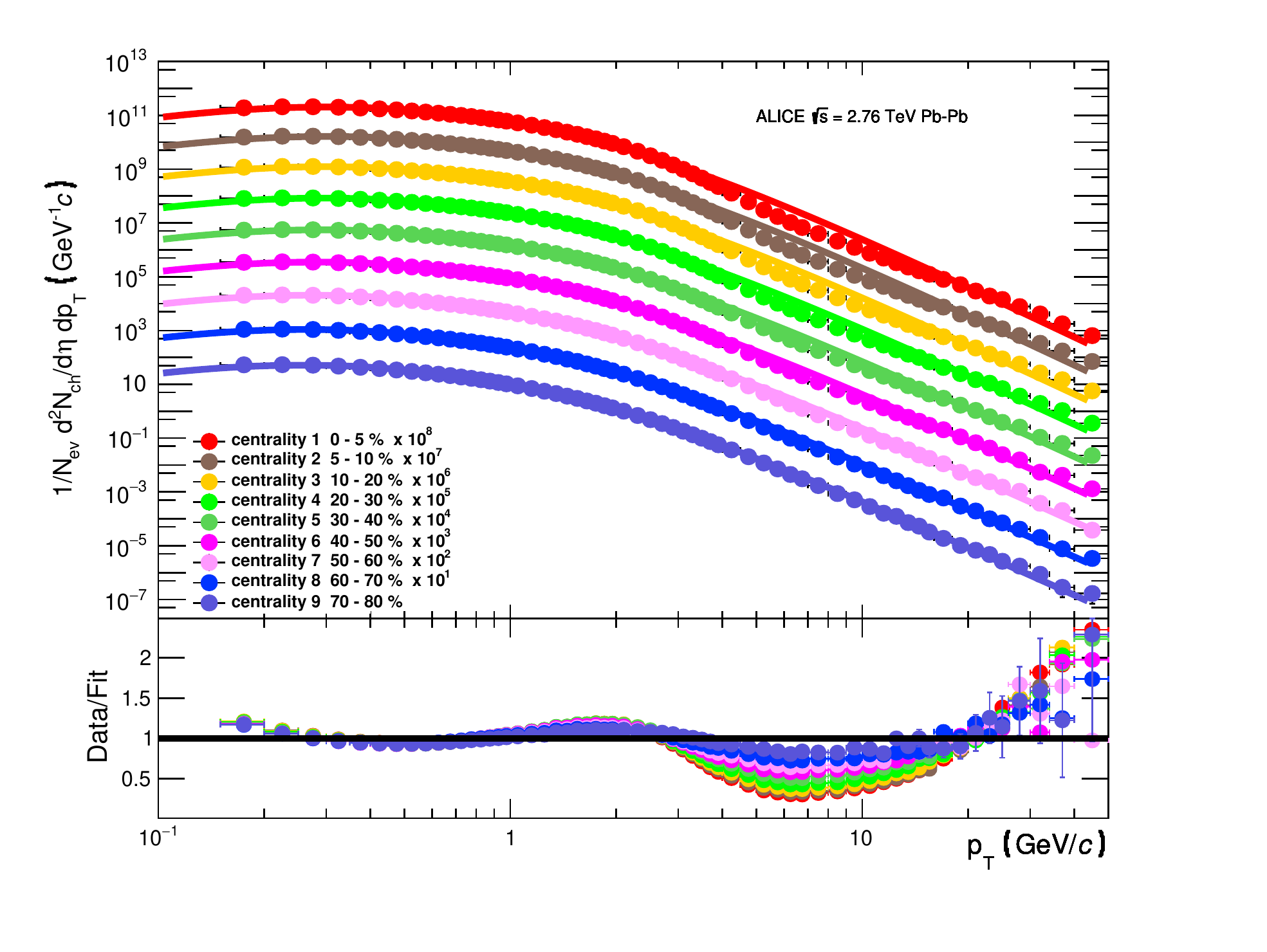}
 \end{minipage}		
\caption{
Transverse momentum distributions of charged hadrons  measured by the 
ALICE collaboration in Pb-Pb collisions at $\sqrt{s_{NN}}$ = 2.76 TeV~\cite{Acharya:2018qsh} for different centrality classes.
The solid lines represent fits made using the Tsallis distribution equation~\eref{tsallisfitcharged} .
The lower part of the figure shows the ratio of the data divided by the fit value.
}
\label{figPbPb276}
\end{figure}

The resulting fits to the experimental data obtained in 
Pb-Pb collisions at $\sqrt{s_{NN}}$ = 2.76 TeV~\cite{Acharya:2018qsh} 
are shown in~\fref{figPbPb276} where we follow the centrality 
classification introduced in~\cite{Acharya:2018qsh}.
As can be seen in figure 1 the fits are very good for  peripheral events and at low $p_T$, 
gradually worsening for the more central events where
the fits at first overshoot the data above $p_T$ values of about 3 GeV  then rejoin the data and  at larger values of 
$p_T$ above about 20 GeV are below
the data.  The same behaviour can be seen for a beam energy of 5.02 TeV in figure 2. We have checked that the 
same behaviour is  also present  in Xe-Xe collisions~\cite{Acharya:2018eaq} at 5.44 TeV. 
The transverse momentum distributions tend to show an S shape for central collisions, this shape is difficult to reproduce using the Tsallis 
parameterization which has only two variables $T$ and $q$ and the overall normalization fixed by the volume $V$. Clearly one more parameter
would be needed to reproduce the shape for the most central events.\\

In \tref{table1} we collect all the results 
for the temperature $T$ and the Tsallis parameter $q$
obtained from fitting the Pb-Pb data at a beam energy of 2.76 TeV. Note that the temperature varies from just slightly
above 96 MeV for the most central events and  to about 78 MeV for the most peripheral events. 
The results obtained this way are consistent with those obtained in analyses using the blast-wave~\cite{Schnedermann:1993ws} 
formalism~\cite{Abelev:2008ab,Chatterjee:2015fua,Retiere:2003kf} but they are considerably lower than those obtained recently 
in~\cite{Prorok:2018okq, Motornenko:2019jha} also the dependence in centrality is reversed. 
It is to be noted 
that fits based on the blast-wave model are based on exponentials and never describe 
data at larger $p_T$ for the simple reason that at large $p_T$ the measured distributions are polynomial and not exponential.
As usual, the Tsallis parameter $q$  can be determined with an excellent accuracy.

\begin{table}[h]
\caption{\label{table1}Values of $q$, $T$ and $\chi^2$/NDF obtained using equation~\eref{tsallisfit} to fit 
the charged hadron transverse momentum spectra data at $\sqrt{s_{NN}}$ = 2.76 TeV~\cite{Acharya:2018qsh}.}
\begin{indented}
\item[]\begin{tabular}{@{}llll}
\br
{\bf Centrality Class} & {\bf$q$} & {\bf$T$ (MeV)}& $\chi^2$/NDF \\
\hline\hline
			1~~~~(0-5)\%     &   1.1355 $\pm$ 0.0009    & 95.9 $\pm$ 1.4  & 156.5/58 \\
			2~~~(5-10)\%     &   1.1363 $\pm$ 0.0009   & 95.5 $\pm$ 1.3  & 150.4/58 \\
			3~~(10-20)\%     &   1.1376 $\pm$ 0.0009   & 94.5 $\pm$ 1.3  & 137.9/58 \\
			4~~(20-30)\%     &   1.1387 $\pm$ 0.0009   & 92.9 $\pm$  1.3  & 117.3/58   \\
			5~~(30-40)\%     &   1.1389 $\pm$ 0.0009   & 91.2 $\pm$  1.3  & 91.47/58  \\
			6~~(40-50)\%     &   1.1403 $\pm$ 0.0009   & 88.0 $\pm$  1.3  & 71.39/58  \\
			7~~(50-60)\%     &   1.1416 $\pm$  0.0010    & 84.6 $\pm$  1.3  & 52.88/58 \\
			8~~(60-70)\%     &   1.1424 $\pm$ 0.0010    & 81.0 $\pm$  1.3  & 29.8/58  \\
			9~~(70-80)\%     &   1.1428 $\pm$ 0.0012    & 78.0 $\pm$  1.3  & 23.16/58 \\
\br
\end{tabular}
\end{indented}
\end{table}

\begin{figure}[ht]
\begin{minipage}{\columnwidth}
\centering
\includegraphics[width=\columnwidth, height = 12.0cm]{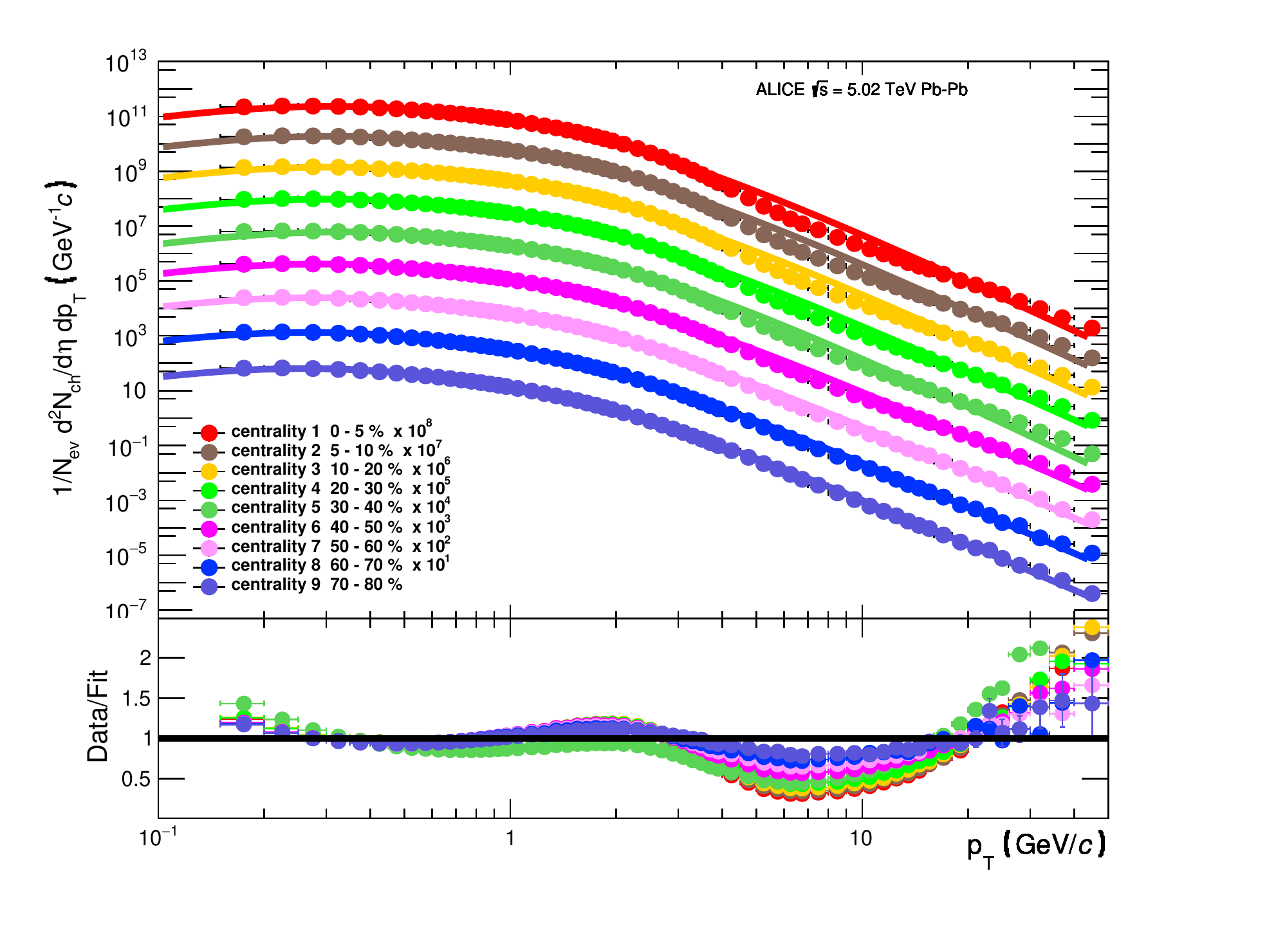}
 \end{minipage}		
\caption{
Transverse momentum distributions of charged hadrons  measured by the 
ALICE collaboration in Pb-Pb collisions at $\sqrt{s_{NN}}$ = 5.02 TeV~\cite{Acharya:2018qsh} for different centrality classes.
The solid lines are fits using the Tsallis distribution~\eref{tsallisfitcharged}.  
The lower part of the figure shows the ratio of the data divided by the fit value.
}
\label{figPbPb502}
\end{figure}

The fits to the experimental data at 5.02 TeV are shown in~\fref{figPbPb502} where, as before, we follow the centrality 
classification from~\cite{Acharya:2018qsh}.
Again the fits are very good for  peripheral events, gradually worsening for the more central events where
the fits at first overshoot the data above $p_T$ values of 2 GeV;  then gradually rejoin the data at larger values of 
$p_T$ and in the end undershoot
the data.

Similar to the procedure followed above, we collect in~\tref{table2}  the results 
for the temperature $T$ and the Tsallis parameter $q$
obtained from fitting the Pb-Pb data at $\sqrt{s_{NN}}$ = 5.02 TeV. 
As in the previous case one has a
very good description of the transverse momentum distributions for the most peripheral collisions, again gradually worsening for the most 
central events where the fits at first overshoot the data at large values of $p_T$ and in the end are below the data. 
The temperature $T$ and the Tsallis parameter $q$ have been determined at the two  beam energies for all the centrality classes. 
\begin{table}
\caption{\label{table2}Values of  $q$, $T$ and $\chi^2$/NDF obtained  using equation~\eref{tsallisfit} to fit the
charged hadron transverse momentum spectra data 
at $\sqrt{s_{NN}}$ =  5.02 TeV~\cite{Acharya:2018qsh}.}
\begin{indented}
\item[]\begin{tabular}{@{}llll} 
\br
{\bf Centrality Class} & {\bf$q$} & {\bf$T$ (MeV)} &$\chi^2$/NDF \\
\hline\hline
			1 ~~~~(0-5)\%     &   1.1405 $\pm$ 0.0009    & 98.2 $\pm$ 1.3 & 163.8/58 \\
			2 ~~~(5-10)\%     &   1.1413 $\pm$ 0.0009     & 97.8 $\pm$ 1.4 & 154.1/58 \\
			3 ~~(10-20)\%     &   1.1424 $\pm$ 0.0009     & 96.8 $\pm$ 1.3 & 142.7/58 \\
			4 ~~(20-30)\%     &   1.1438 $\pm$ 0.0009    & 94.8 $\pm$ 1.2 & 126.6/58   \\
			5 ~~(30-40)\%     &   1.1449  $\pm$  0.0009  & 92.5 $\pm$ 1.2 & 104.9/58  \\
			6 ~~(40-50)\%     &   1.1467 $\pm$ 0.0009    & 88.8  $\pm$ 1.2 & 86.17/58  \\
			7 ~~(50-60)\%     &   1.1478 $\pm$ 0.0009    & 85.3  $\pm$ 1.2 & 61.57/58  \\
			8 ~~(60-70)\%     &   1.1489 $\pm$ 0.0009    & 81.3 $\pm$ 1.2 & 37.62/58  \\
			9 ~~(70-80)\%     &   1.1503 $\pm$ 0.0010   & 77.4 $\pm$ 1.2  & 30.3/58 \\
\br
\end{tabular}
\end{indented}
\end{table}

\section{Thermodynamic Variables}
\subsection{Energy Density at Kinetic Freeze-Out}
Having deduced the temperature $T$ and the Tsallis parameter $q$ at kinetic freeze-out from the transverse momentum
distributions for two beam energies, we now proceed calculating the energy density given by equation~\eref{epsilon}:
\begin{equation}
\epsilon = 2\sum_{i=1}^3 g_i \int\frac{d^3p}{(2\pi)^3}E_i\left(1+(q-1)\frac{E_i}{T}\right)^{-\frac{q}{q-1}},
\end{equation}
where $i = \pi^+, K^+, p$. 
As before, the factor 2 on the right hand side takes into account the contributions 
from antiparticles, $\pi^-, K^-$ and $\bar{p}$. 
The results are shown in~\tref{tablee} as a function of centrality
and compare with a few other energy densities. 
The entry for the chemical freeze-out energy density has been obtained using the latest version of 
THERMUS~\cite{Wheaton:2004qb}~\footnote{B. Hippolyte and Y. Schutz, https://github.com/thermus-project/THERMUS}.  
The latter has been calculated from all hadronic resonances and is not limited to the
charged particles only.  
It has been shown recently that the chemical freeze-out temperature is approximately 
independent of centrality~\cite{Abelev:2008ab,Sharma:2018jqf,Vovchenko:2019kes}.
For comparison we also show 
the energy density inside a proton  calculated using the charge radius of the proton given as 0.875 fm and the mass of 
the proton as listed in the PDG~\cite{Tanabashi:2018oca}.
The difference between the kinetic and chemical freeze-out results is not surprising 
in view of the fact that the energy density  changes as $T^4$ for massless particles.  
In~\tref{tablee} we also show  the energy density obtained in the phase transition region obtained using Lattice QCD~\cite{Ding:2015ona}.
 

 
In figure 3 we show the  energy density divided by the kinetic freeze-out temperature to the fourth power
so as to have a dimensionless quantity. As can be seen in figure,  in this case the dependence on the centrality
 class is strongly reduced.
 
\begin{figure}[ht]
\begin{minipage}{\columnwidth}
\centering
\includegraphics[width=\columnwidth, height = 9.0cm]{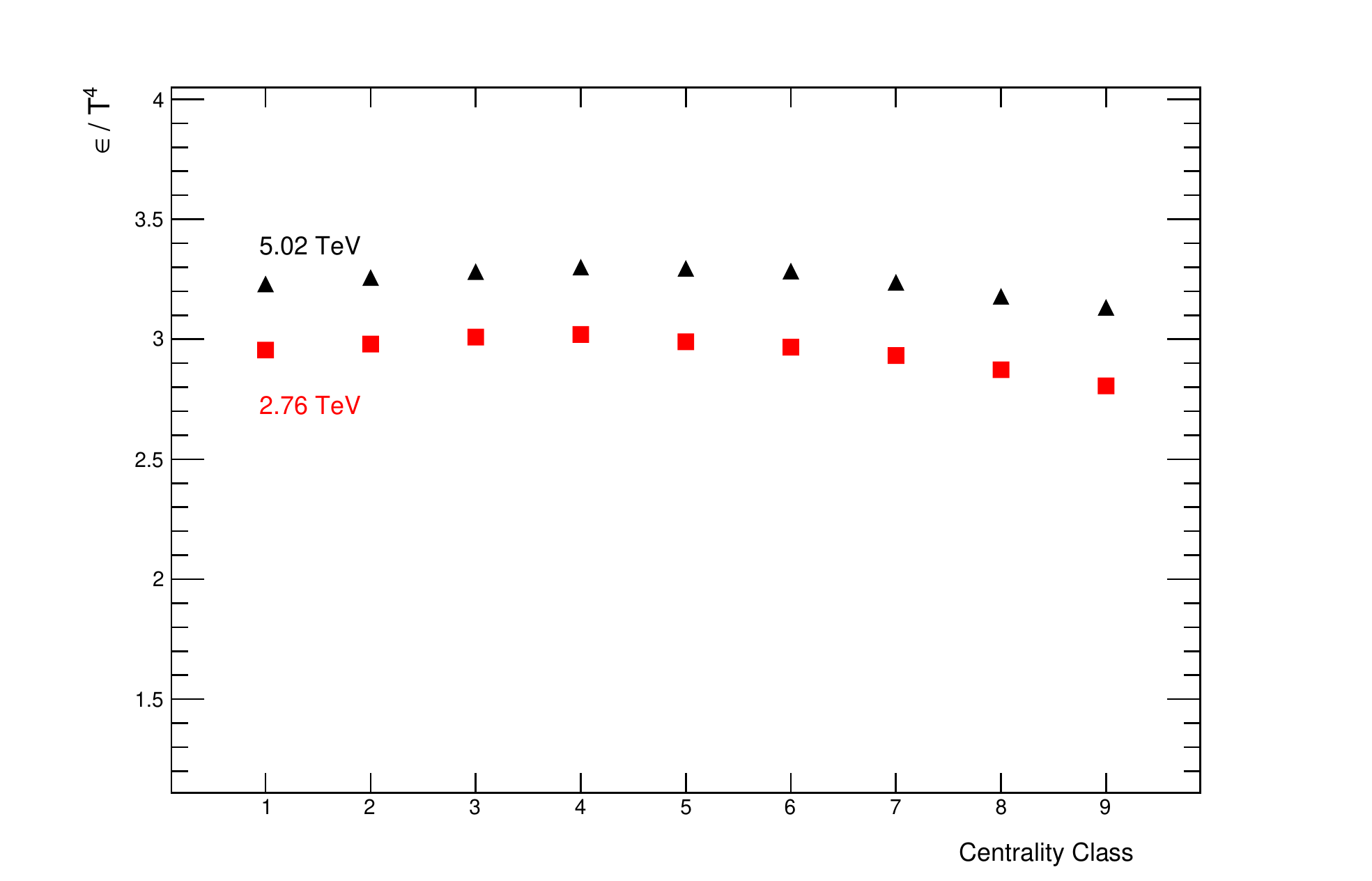}
 \end{minipage}
\caption{
Energy density of charged hadrons divided by the kinetic freeze-out temperature
 at kinetic freeze-out in Pb-Pb collisions at 2.76 and 5.02  TeV~\cite{Acharya:2018qsh} as a function of
centrality class calculated using equation~\eref{epsilon}.
}

\label{fig_epsilon_scaled}
\end{figure}
 
\begin{table}
\caption{\label{tablee} Values for the energy density of charged hadrons, 
expressed in GeV/fm$^3$, obtained using equation~\eref{epsilon} for 
the different centrality classes. The energy density at chemical freeze-out has been
calculated at $T$ = 153 $\pm$ 3 MeV for the most central Pb-Pb  collisions as given in~\cite{Sharma:2018jqf}.}
\begin{indented}
\item[]\begin{tabular}{@{}lll}
\br
{\bf Centrality Class} & {\bf$\epsilon$} at 2.76 TeV&   {\bf$\epsilon$} at 5.02 TeV\\
\hline\hline
			1~~~~(0-5)\%      &0.03272 $\pm$ 0.00041        & 0.03933 $\pm$    0.00049        	\\
			2~~~(5-10)\%      &0.03218 $\pm$ 0.00041       & 0.03860  $\pm$      0.00049     	\\
			3~~(10-20)\%      &0.03153 $\pm$ 0.00039        & 0.03732 $\pm$     0.00047      	\\
			4~~(20-30)\%      &0.02938 $\pm$  0.00037       & 0.03487  $\pm$    0.00045       	\\
			5~~(30-40)\%      &0.02696  $\pm$  0.00035      & 0.03148 $\pm$    0.00042       	\\
			6~~(40-50)\%      &0.02339 $\pm$  0.00032        & 0.02669 $\pm$    0.00036       	\\
			7~~(50-60)\%      &0.01964 $\pm$    0.00028       & 0.02241	$\pm$     0.00031      \\
			8~~(60-70)\%      &0.01604 $\pm$     0.00025      & 0.01809	$\pm$    0.00026       \\
			9~~(70-80)\%      &0.01356  $\pm$     0.00024      & 0.01458	$\pm$    0.00023       \\
\hline
proton~\cite{Tanabashi:2018oca}&  0.334 &\\
chemical freeze-out~\cite{Wheaton:2004qb}&  0.36 $\pm$ 0.07&\\
lattice QCD~\cite{Ding:2015ona}&  0.34 $\pm$ 0.16&\\
cold nuclear matter&  0.16&\\
\br
\end{tabular}
\end{indented}
\end{table}
An estimate can now be made of the lifetime of the hadronic stage between  chemical freeze-out and the kinetic freeze-out using the Bjorken 
model~\cite{Bjorken:1982qr}  with isentropic expansion which gives:
\begin{equation}
\epsilon(\tau) = \epsilon(\tau_0)\left(\frac{\tau_0}{\tau}\right)^{4/3}  .
\end{equation}
For the top 5\% central collisions at 5.02 TeV this leads to
\begin{equation}
\frac{\tau({\mathrm{kinetic~~fo}})}{\tau({\mathrm{chemical~~fo})}} 
= \left(\frac{\epsilon({\mathrm{chemical}})}{\epsilon({\mathrm{kinetic}})}\right)^{3/4}\approx 3.9  ,
\end{equation}
where the energy density at kinetic freeze-out has been corrected by a factor 3/2 to take into account the neutral hadrons.
i.e. if chemical freeze-out happens at  $\tau$ = 10 fm, then kinetic freeze-out happens at $\tau$ = 39 fm. 
The chemical freeze-out time  could be different 
for different centralities.  If the chemical freeze-out time is the same or at least similar for all centralities
then one has to conclude that the time between chemical and kinetic freeze-out is longer for peripheral collisions
than for central collisions.
As a reminder, 
in the Bjorken model~\cite{Bjorken:1982qr}, which is an inside-outside cascade, the central region freezes out first while the 
peripheral region remains hot.  As this is a scaling model, there is no natural cut-off time.  
\subsection{Pressure at Kinetic Freeze-Out}
The pressure plays an important role in the hydrodynamic description of heavy-ion collisions, e.g. in the study of shock waves
or the speed of sound in a hadronic gas.
In the present analysis, the pressure can be determined explicitly from the following equation~\eref{pressure}:
\begin{equation}
P = 2 \sum_{i=1}^3 g_i\int\frac{d^3p}{(2\pi)^3}\frac{p^2}{3E_i}\left(1+(q-1)\frac{E_i}{T}\right)^{-\frac{q}{q-1}},
\end{equation}
where $i = \pi^+, K^+, p$. 
As previously, the factor 2 on the right hand side takes into account the contributions 
from antiparticles, $\pi^-, K^-$ and $\bar{p}$. \\
\begin{figure}[ht]
\includegraphics[width=\columnwidth, height = 9.0cm]{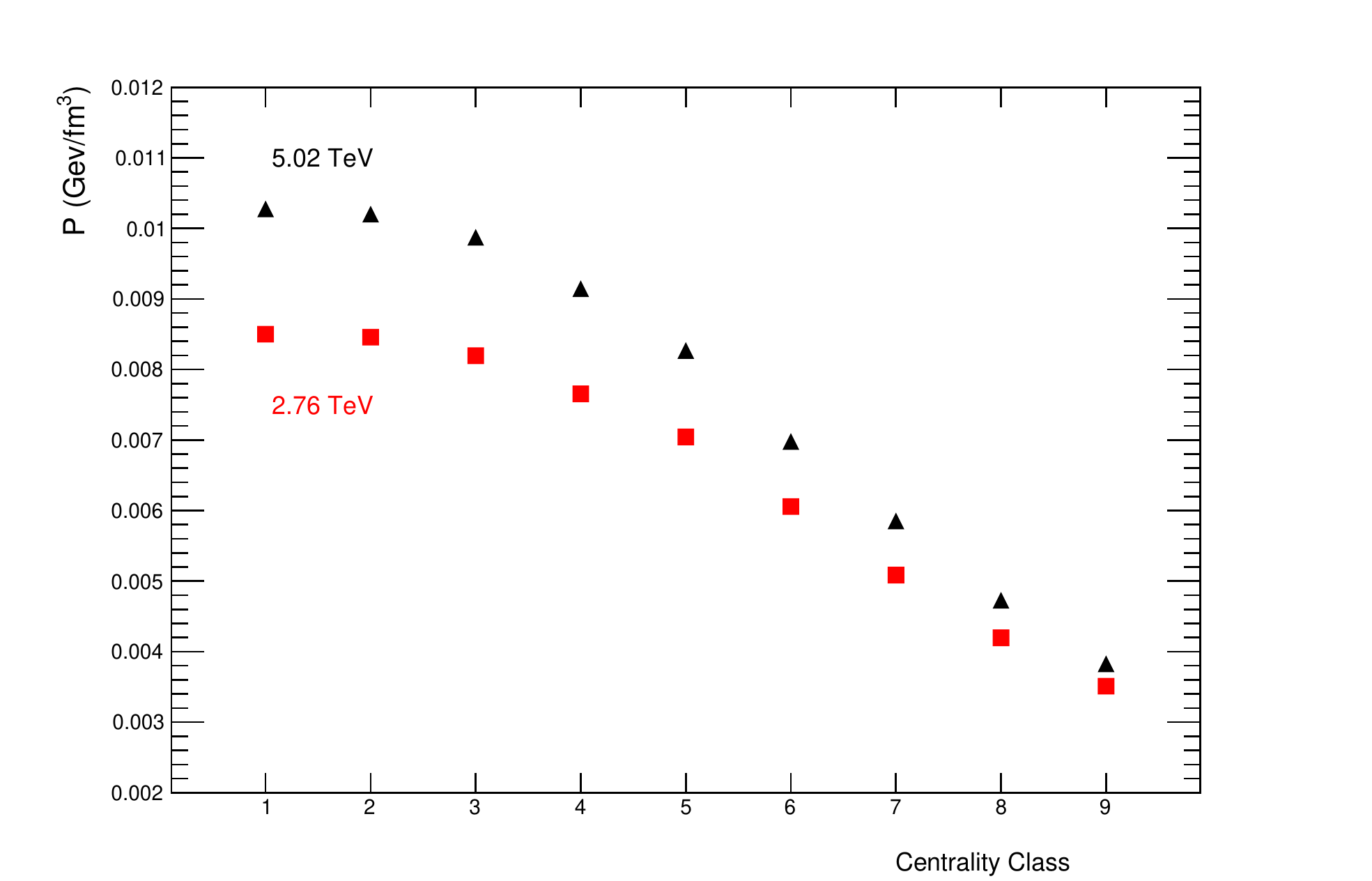}
\caption{
Pressure of charged hadrons at kinetic freeze-out in Pb-Pb collisions at 2.76 and 5.02  TeV~\cite{Acharya:2018qsh} as a function of
centrality class calculated using equation~\eref{pressure}.
}

\label{pressure_276_502}
\end{figure}
The results are shown in~\fref{pressure_276_502} where one notices a clear, expected, increase in the pressure when going from 
peripheral collisions to central ones.
We have also checked explicitly that the inequality:
\begin{equation}
\epsilon \geq 3P  ,
\end{equation}
is always satisfied.

\subsection{Entropy Density at Kinetic Freeze-Out}
\begin{figure}[ht]
\begin{center}
\includegraphics[width=\columnwidth, height = 9.0cm]{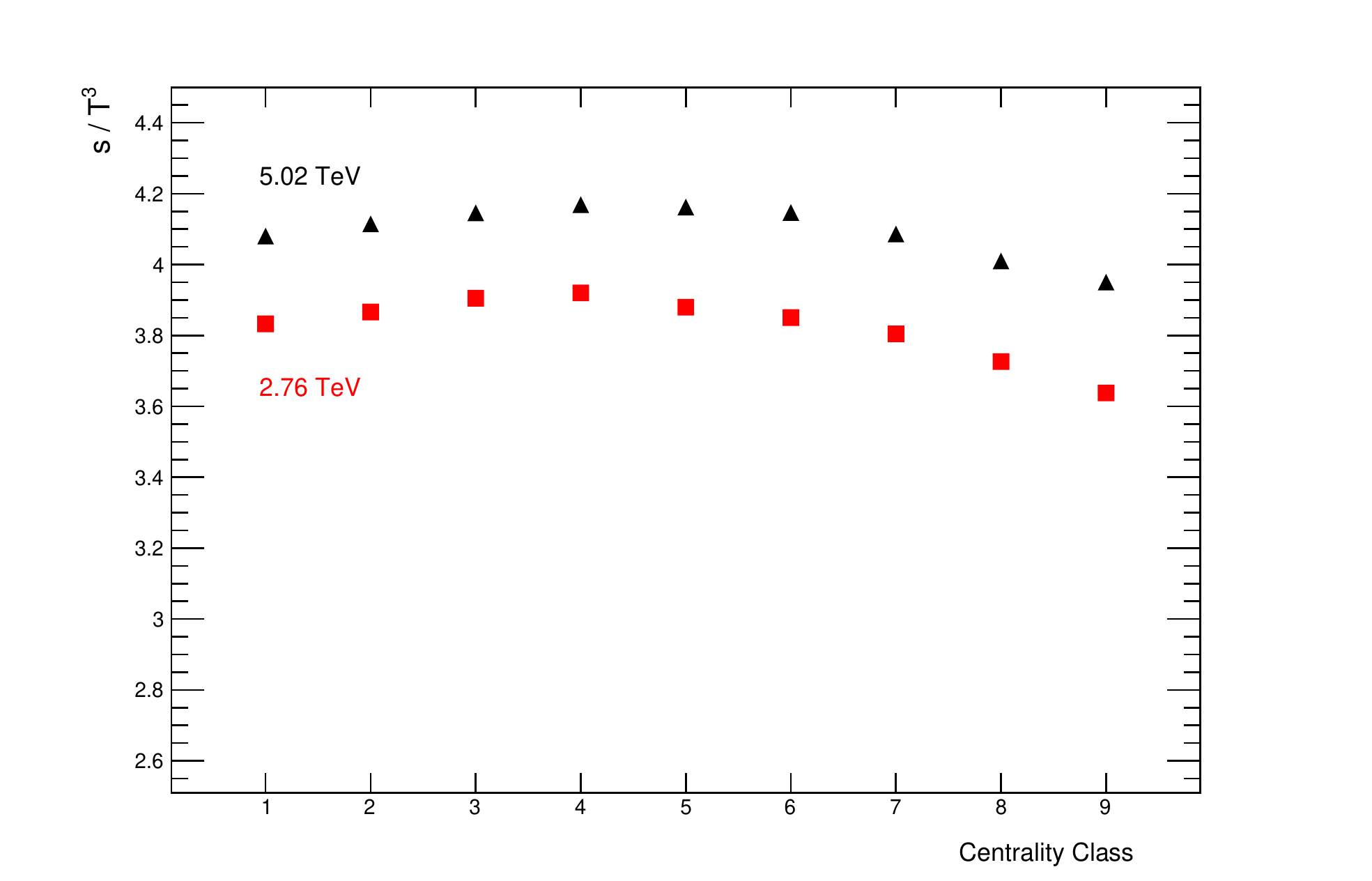}
\caption{
Entropy density divided by the kinetic freeze-out temperature to the third power  of charged hadrons 
 at kinetic freeze-out in Pb-Pb collisions at 2.76 and 5.02  TeV~\cite{Acharya:2018qsh} as a function of
centrality class  calculated using equation~\eref{entropy}   .
}

\label{entropy_276_502}
\end{center}
\end{figure}

The entropy is an important quantity because it plays a major role in hydrodynamic expansion calculations where entropy is sometimes assumed
to be conserved when going from the quark-gluon plasma phase to the hadronic phase. This is for example the case in the Bjorken 
model~\cite{Bjorken:1982qr}.  It is  difficult to relate it directly to a measurable quantity
and it is often indirectly linked to the particle number. In this paper the connection is a fairly direct one and can be obtained using 
equation~\eref{entropy}.
More explicitly, the entropy density is given by the following expression  where the parameters $T$ and $q$ are 
taken from~\tref{table1} for Pb-Pb collisions at $\sqrt{s_{NN}}$ = 2.76 TeV and~\tref{table2} for collisions at 5.02 TeV respectively:
\begin{equation}
\hspace*{-1cm} s =  2 \sum_{i=1}^3 g_i\int\frac{d^3p}{(2\pi)^3}
\left[\frac{E_i}{T}\left(1+(q-1)\frac{E_i}{T}\right)^{-\frac{q}{q-1}} +\left(1+(q-1)\frac{E_i}{T}\right)^{-\frac{1}{q-1}}\right] , 
\end{equation}
where, as before,  $i = \pi^+, K^+, p$. 
The factor 2 on the right hand side, as previously, takes into account the contributions 
from antiparticles, $\pi^-, K^-$ and $\bar{p}$. 
The results are shown in~\fref{entropy_276_502}
 where the entropy density has been divided by $T^3$ so as to have a dimensionless quantity.
  There is also a small increase when the beam energy is increased from  $\sqrt{s_{NN}}$ = 2.76 to 5.02 TeV.
We have also checked explicitly that the thermodynamic relation,
\begin{equation}
\epsilon + P = Ts ,
\end{equation}
holds. This is further confirmation of the consistency of having the chemical potential $\mu$  equal to zero for the collisions under consideration. 
As this is done at kinetic freeze-out and not at chemical freeze-out, this is a non-trivial observation. At chemical freeze-out 
the chemical potentials must be zero because of the equal numbers of particles and antiparticles. At thermal freeze-out however
it is only required that the chemical potentials for particles and antiparticles be equal but not necessarily zero. It is still legitimate to 
have chemical potentials at kinetic freeze-out but they change the normalization and no longer determine relative abundancies.

\begin{figure}[ht]
\begin{center}
\includegraphics[width=\columnwidth, height = 9.0cm]{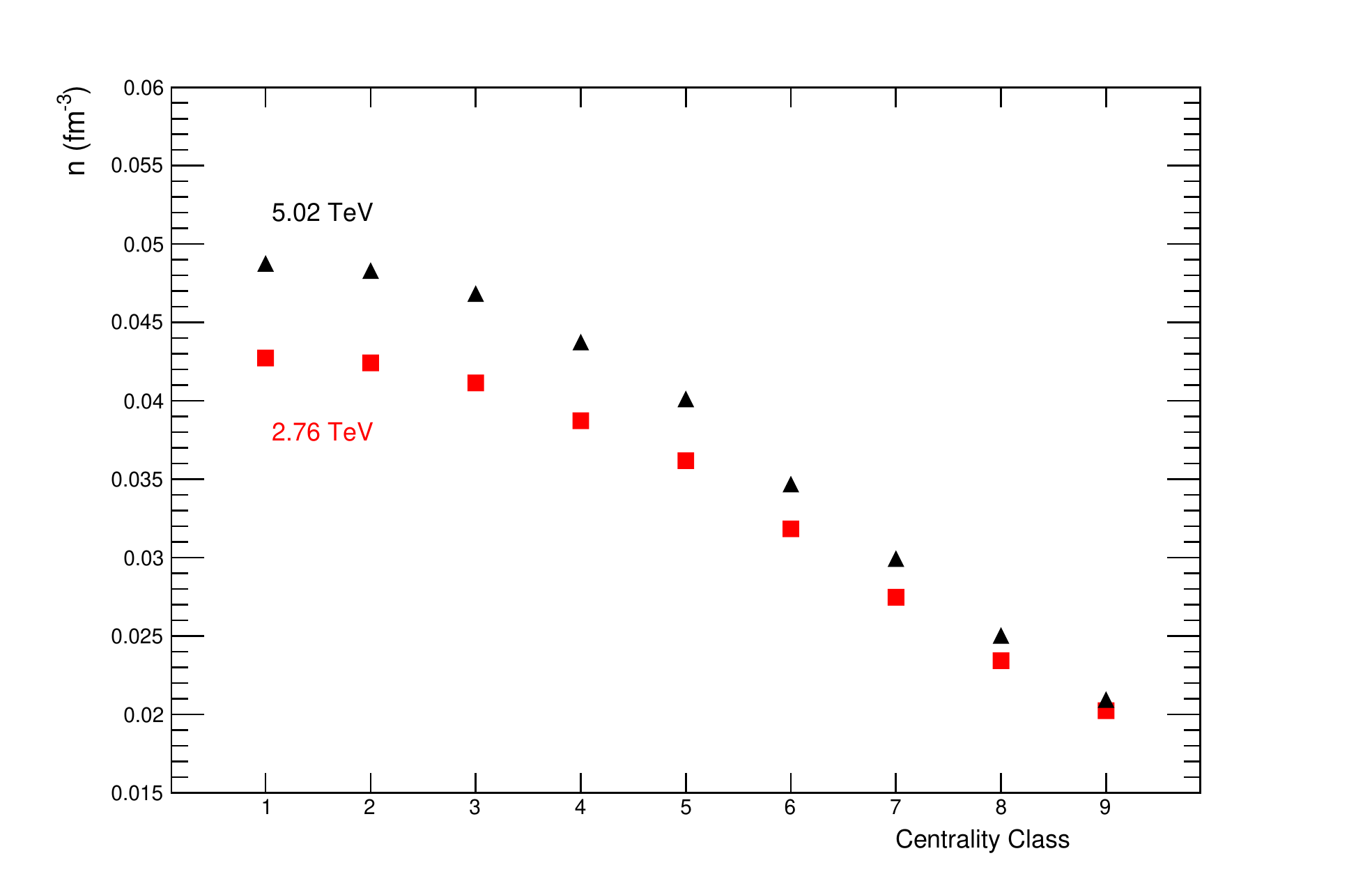}
\caption{
Charged particle density at kinetic freeze-out in Pb-Pb collisions at 2.76 and 5.02  TeV~\cite{Acharya:2018qsh} as a function of
centrality calculated using equation~\eref{Number}   .
}

\label{particle_density_276_502}
\end{center}
\end{figure}

For completeness we show the particle density calculated using equation~\eref{Number} in~\fref{particle_density_276_502}. 
This is clearly well below the 
interior density of a heavy nucleus which is 0.17 nucleons/fm$^3$~\cite{Tanabashi:2018oca}.
\section{Summary}
The transverse momentum distributions of the primary charged particles measured in Pb - Pb collisions at $\sqrt{s_{NN}}$ = 2.76 and 5.02 TeV
 by the ALICE collaboration~\cite{Acharya:2018qsh} 
have been analysed in this paper using a thermodynamically consistent form of the Tsallis distribution based on equation~\eref{tsd1}.  
This gives a
very good description of the transverse momentum distributions for the most peripheral collisions, gradually worsening for the most 
central events where the fits at first overshoot the data at large values of $p_T$ and in the end are below the data, which is a matter of further exploration. 
The temperature $T$ and the Tsallis parameter $q$ have been determined at the two  beam energies for all the centrality classes. Using the values obtained 
we then determined 
 the energy density, $\epsilon$, pressure, $P$, entropy density, $s$ and the particle density, $n$ at kinetic freeze-out 
 as a function of the  centrality classes. As expected, the  values of all the thermodynamic quantities
 show an increase towards higher centrality class and at higher beam energy.
It is determined that in the final freeze-out stage the energy density
reaches a value of about 0.039 GeV/fm$^3$ for the most central collisions at $\sqrt{s_{NN}}$ =  5.02 TeV. This is less than that at chemical
freeze-out 
where the energy density is about 0.36 GeV/fm$^3$. This decrease approximately follows a $T^4$ law. It can be concluded that, together with the results
obtained at chemical freeze-out, the thermodynamic quantities presented in this paper provide information about the evolution
of the thermodynamic quantities during the evolution of the hadronic phase from chemical to kinetic freeze-out.

\ack
One of us (T.B.) acknowledges  partial support from the joint research projects between the JINR and IFIN-HH.
We thank Smbat Grigoryan for  helpful comments.


\newpage
\section*{References}

\begin{thebibliography}{30}
\bibitem{Acharya:2018qsh}  Acharya S et al.  2018 \JHEP  {\bf 11} 013
\bibitem{Tsallis:1987eu} Tsallis C 1988 {\sl J. Statist. Phys.} {\bf 52} 479 1988
\bibitem{Abelev:2013ala}  Abelev B B et al. 2013 {\sl Eur. Phys. J.}  C {\bf 73} 2662
\bibitem{Biyajima:2006mv} Biyajima M,  Mizoguchi T,  Nakajima N,  Suzuki N, and Wilk G 2006 {\sl Eur. Phys. J.} C {\bf 48} 597
\bibitem{Lao:2016gxv} Lao H-L, Liu F-H and Lacey R A  2017 {\sl Eur. Phys. J.} A {\bf 53} 44 [Erratum: 2017 {\sl Eur. Phys. J.}  A {\bf 53}  13]
\bibitem{Si:2017cyg} Si R-F Li H-L and Liu F-H  2018 {\sl Adv. High Energy Phys.}  {\bf 2018} 7895967
\bibitem{Biro:2017arf} B\'{ı}r\'{o} G, Barnaf\"{o}ldi G G, Bir\'{o} T S,  \"{U}rm\"{o}ssy K and  Tak\'{a}cs \'{A} 2017 {\sl Entropy}  {\bf 19} 88
\bibitem{Hui:2017zqy} Hui J-Q, Jiang Z-J and  Xu D-F 2018 {\sl Adv. High Energy Phys.} {\bf 2018} 7682325
\bibitem{Cleymans:2011in} Cleymans J and  Worku D 2012 {\sl Eur. Phys. J.}  A{\bf 48} 160
\bibitem{Cleymans:2012ya} Cleymans J and  Worku D 2012 \jpg {\bf 39} 025006
\bibitem{Wong:2013sca} Wong C-Y and Wilk G 2013 \PR  D {\bf 87} 114007
\bibitem{Wong:2015mba}  Wong C-Y, Wilk G, Cirto L J L  and Tsallis C   2015 \PR  D {\bf 91} 114027 
\bibitem{Azmi:2014dwa} Azmi M D  and Cleymans J 2014 \jpg {\bf 41} 065001 
\bibitem{Azmi:2015xqa} Azmi M D and Cleymans J 2015 {\sl Eur. Phys. J.}  C {\bf 75} 430
\bibitem{Grigoryan:2017gcg} Grigoryan S 2017 \PR  D {\bf 95} 056021
\bibitem{Acharya:2018eaq}  Acharya  S et al. 2019 \PL B {\bf 788} 166
\bibitem{Schnedermann:1993ws}  Schnedermann E, Sollfrank J and Heinz U W 1993 \PR  C {\bf 48} 2462
\bibitem{Abelev:2008ab} Abelev B I at al. 2008 \PR C {\bf 79} 034909 
\bibitem{Chatterjee:2015fua} Chatterjee S, Das S, Kumar L, Mishra D, Mohanty B, Sahoo R and  Sharma N 2015 {\sl Adv. High
 Energy Phys.}  {\bf 2015} 349013
\bibitem{Retiere:2003kf} Retiere F and Lisa M A 2004 \PR  C {\bf 70} 044907
\bibitem{Prorok:2018okq} Prorok D  2019 {\sl Eur. Phys. J.}  A {\bf 55} 37
\bibitem{Motornenko:2019jha} Motornenko A,  Vovchenko V, Greiner C and Stoecker H  2019 arXiv:1908.11730 [hep-ph] 
\bibitem{Tanabashi:2018oca} Tanabashi M et al. 2018 \PR  D {\bf 98} 030001
\bibitem{Wheaton:2004qb}  Wheaton S, Cleymans J and Hauer M 2009 {\sl Comput. Phys. Commun.}  {\bf 180} 84
\bibitem{Ding:2015ona} Ding H-T, Karsch F, and Mukherjee S  2015 {\sl Int. J. Mod. Phys.}  E {\bf 24} 1530007
\bibitem{Sharma:2018jqf} Sharma N, Cleymans J, Hippolyte B and Paradza M 2019 \PR C {\bf 99} 044914 
\bibitem{Vovchenko:2019kes}  Vovchenko V, D\"{o}nigus B and Stoecker H 2019 arXiv:1906.03145 [hep-ph] 
\bibitem{Bjorken:1982qr} Bjorken J D  1983 \PR  D {\bf 27} 140
\end{thebibliography}
\end{document}